%% file: main.tex
\documentclass[a4paper,twocolumn,11pt,accepted=2025-12-01]{quantumarticle}
\pdfoutput=1
\usepackage[utf8]{inputenc}
\usepackage{xcolor}
\usepackage[english]{babel}
\usepackage[T1]{fontenc}
\usepackage[numbers,sort&compress]{natbib}
\usepackage{amsmath,amssymb}
\usepackage{braket}
\usepackage[ruled]{algorithm2e}
\usepackage{hyperref}
\usepackage{ulem}

\newcommand{\QUITS}{\texttt{QUITS}}

\newcommand{\dqc}{Duke Quantum Center, Duke University, Durham, NC 27701, USA}
\newcommand{\physics}{Department of Physics, Duke University, Durham, NC 27708, USA}
\newcommand{\ece}{Department of Electrical and Computer Engineering, Duke University, Durham, NC 27708, USA}

\usepackage{caption,subcaption,graphicx}
\usepackage{tikz,pgfplots,pgf}
\usepackage{xcolor}

\begin{document}

\title{QUITS: A modular Qldpc code circUIT Simulator}

\author{Mingyu Kang}
\affiliation{\dqc}
\affiliation{\physics}
\email{mingyu.kang@duke.edu}
\thanks{These authors contributed equally.}

\author{Yingjia Lin}
\affiliation{\dqc}
\affiliation{\physics}
\email{yingjia.lin@duke.edu}
\thanks{These authors contributed equally.}

\author{Hanwen Yao}
\affiliation{\dqc}
\affiliation{\ece}

\author{Mert G\"okduman}
\affiliation{\dqc}
\affiliation{\ece}

\author{Arianna Meinking}
\affiliation{\dqc}
\affiliation{\physics}

\author{Kenneth R. Brown}
\affiliation{\dqc}
\affiliation{\physics}
\affiliation{\ece}
\affiliation{Department of Chemistry, Duke University, Durham, NC 27708, USA}

\maketitle

\begin{abstract}
\noindent
To achieve quantum fault tolerance with lower overhead, quantum low-density parity-check (QLDPC) codes have emerged as a promising alternative to topological codes such as the surface code, offering higher code rates. To support their study, an end-to-end framework for simulating QLDPC codes at the circuit level is needed. In this work, we present \QUITS, a modular and flexible circuit-level simulator for QLDPC codes. Its design allows users to freely combine LDPC code constructions, syndrome extraction circuits, decoding algorithms, and noise models, enabling comprehensive and customizable studies of the performance of QLDPC codes under circuit-level noise. \QUITS\ supports several leading QLDPC families, including hypergraph product codes, lifted product codes, and balanced product codes. As part of the framework, we introduce a syndrome extraction circuit improved from Tremblay, Delfosse, and Beverland [Phys. Rev. Lett. 129, 050504 (2022)] that applies to all three code families. In particular, for a small hypergraph product code, our circuit achieves lower depth than the conventional method, resulting in improved logical performance. Using \QUITS, we evaluate the performance of state-of-the-art QLDPC codes and decoders under various settings, revealing trade-offs between the decoding runtime and the logical failure rate. The source code of \texttt{QUITS} is available online~\footnote{\url{https://github.com/mkangquantum/quits}}.
\end{abstract}

\section{Introduction}

Quantum error correction (QEC) is a key requirement for building a scalable and fault-tolerant quantum computer. The theory of quantum fault tolerance establishes that the encoded quantum information can be protected even in the presence of \textit{circuit-level noise}, a noise model in which errors may occur at any qubit, gate, and measurement involved in the QEC~\cite{gottesman1998theory, aharonov1997fault}. 

\begin{figure}[ht!]
  \centering
  \includegraphics[width=0.92\linewidth]{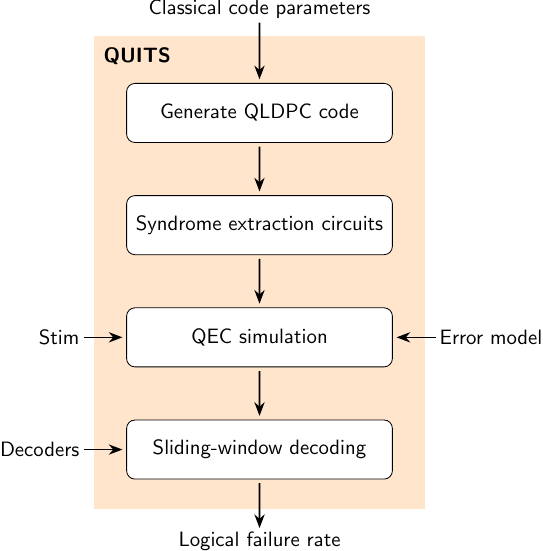}
  \caption{\textbf{Overview of \QUITS.} QUITS provides a streamlined framework for QLDPC codes simulation. The shaded areas are the modularized functions of QUITS. Given code parameters and an error model as the input, QUITS calls Stim and customized decoders to perform QLDPC simulation and estimates the logical failure rate. } 
  \label{fig:QUITS}
\end{figure}

The QEC performance heavily depends on factors such as the choice of the QEC code, the design of the circuit, and the implementation of the decoder. As such, a unified and modular framework is highly desired for evaluating the performance of each component. 

Recently, research in QEC has been facilitated by software packages such as the stabilizer circuit simulator Stim~\cite{gidney2021stim} and the minimum-weight perfect matching decoder PyMatching~\cite{higgott2022pymatching}. For surface codes~\cite{fowler2012surface} in particular, Stim and PyMatching together offer a versatile end-to-end framework for simulating the circuit-level performance of surface codes~\cite{higgott2023improved, wu2023fusion, skoric2023parallel, cain2024correlated, sahay2025error,
higgott2025sparse}. For higher rate codes such as quantum low-density parity-check (QLDPC) codes with non-local connectivity~\cite{breuckmann2021quantum}, circuit-level simulations have been performed in various works~\cite{tremblay2022constant, xu2024constant, bravyi2024high} using Stim; however, open-source frameworks that enable end-to-end simulations of the circuit are not widely developed, with the notable exception of Ref.~\cite{gong2024toward} specifically for bivariate bicycle codes~\cite{bravyi2024high}. 

Even when the QLDPC code and the decoder are given, several challenges in performing circuit-level simulations remain. First, finding a circuit for extracting the syndromes of a QEC code in parallel is not straightforward from the code's parity-check matrices. Previous works often rely on a brute-force search to find the optimal scheduling of the entangling gates in the syndrome extraction circuit~\cite{beverland2021cost, bravyi2024high, pato2024concatenated}. Second, decoders for QLDPC codes are often designed for the code-capacity noise model, where errors in gates and measurements are ignored. Additional work needs to be done to use the decoders for handling the correlated errors at the circuit level. 

This work presents \QUITS, a modular circuit simulator for QLDPC codes. Fig.~\ref{fig:QUITS} provides an overview of our software package and how it can be interfaced with other existing packages such as Stim~\cite{gidney2021stim} and LDPC decoders~\cite{Roffe_LDPC_Python_tools_2022, roffe2020decoding}. Our simulator addresses the challenges in designing syndrome extraction circuits and implementing decoders for simulating QLDPC codes at the circuit level. 

First, we provide a framework for writing the parallelized syndrome extraction circuit of QLDPC codes obtained from (generalized) products of two classical codes. Specifically, we improve the construction of the cardinal circuit, originally proposed for hypergraph product codes~\cite{tremblay2022constant}, and generalize to quasi-cyclic lifted product codes~\cite{panteleev2021degenerate} and balanced product cyclic codes~\cite{tiew2024low}. We expect our simulator to be helpful in finding the syndrome extraction circuit for newly developed QLDPC codes with a similar structure. 

Second, we provide a modular framework for implementing decoders at the circuit level. In this framework, syndromes are divided into overlapping windows and passed to an \textit{inner decoder} along with circuit-level noise information obtained from Stim \cite{gidney2021stim}, which then suggests a correction for each decoding window. The modular structure of this framework enables easy replacement of inner decoders, while the flexibility to adjust decoding window sizes mitigates the reduction in effective distance caused by the finite decoding window. Additionally, the conversion function from the Stim detector error model to window slices of the detector error matrix facilitates the analysis of various circuit-level noise models. As a result, this framework streamlines the development and testing of different decoders under circuit-level noise conditions.

This paper is organized as follows. In Sec.~\ref{sec:background}, we review some basics of QEC and QLDPC codes. In Sec.~\ref{sec:qeccodes}, we introduce the QLDPC codes considered in this work. We review the syndrome extraction circuit in Ref.~\cite{tremblay2022constant} and discuss our improvements
in Sec.~\ref{sec:syndcircuits}. We introduce our decoding framework with a focus on its modular structure in Sec.~\ref{sec:decoder}. Finally, we present our numerical results in Sec.~\ref{sec:results} and conclude the paper with outlooks in Sec.~\ref{sec:outlook}.

\section{Background}\label{sec:background}

Stabilizer codes of $n$ physical qubits encode $k$ logical qubits in the $+1$ eigenspace of an Abelian group $S$ of Pauli operators, the stabilizer group. The set of logical Pauli operators is all Pauli operators that commute with $S$ but are not in $S$. The weight of a Pauli operator is defined as the number of physical qubits it acts on non-trivially. The distance $d$ of a stabilizer code is the minimum weight of a logical Pauli operator. We denote a stabilizer code by the set of code parameters $[[n,k,d]]$. 

A set of generators of $S$, or \textit{check} operators, is measured at each cycle of QEC, where each measurement result is a \textit{syndrome}. The minimum set of generators contains $n-k$ independent generators. Errors that bring the state outside the codespace flip some of the check operators and fire the corresponding syndromes. The \textit{decoder} takes the syndromes as input and returns a suggested \textit{correction} that reproduces the same syndrome. Applying this correction maps the state back to the codespace. If the correction and the error form a stabilizer, the code state is restored. Otherwise, they form a logical operator and create a logical error. Typically, a minimum-weight decoder is used, which chooses the correction with the lowest possible weight, or the error combination that occurs with the highest probability, that yields the same syndrome \cite{dennis2002topological}. With a perfect minimum-weight decoder, up to $\lfloor (d-1)/2 \rfloor$ errors can be reliably corrected.

Calderbank-Shor-Steane (CSS) codes  \cite{calderbank1996good,steane1996multiple} are stabilizer codes with a generator set containing either only Pauli-$Z$ operators or only Pauli-$X$ operators. We can represent the $P$-type ($P=Z,X$) checks using the parity check matrix $H_P$. The elements of the matrix are given by $H_P^{ij}=1$ only if the $i$th $P$-type check has support on the $j$th qubit; otherwise, $H_P^{ij}=0$. The check matrices satisfy the relation $H_XH_Z^{T}=0$ such that all the checks commute. Measuring $Z$($X$)-type checks can detect $X$($Z$)-type errors that anticommute with the check operator. Thus, the $X$ and $Z$-type errors can be decoded independently.

The \textit{Tanner graph}, a bipartite graph that represents the parity-check matrix of a classical or quantum code, is constructed as follows. One set of vertices represents the bits or qubits that encode the data, and the other set of vertices represents the checks. Edges connect each check vertex to each vertex that represents the bit or qubit that supports the check. 

The circuit implementation of QEC involves \textit{data qubits}, which contain the encoded information, and \textit{syndrome qubits}, which are measured to reveal the syndromes. At each QEC cycle, each check operator is transferred to the corresponding syndrome qubit via \textit{syndrome extraction circuit}, which includes multiple layers of two-qubit gates. The ordering of the two-qubit gates needs to be carefully designed to correctly extract the syndromes in parallel. QEC should be capable of protecting quantum information under the \textit{circuit-level noise model}, where errors may occur at any qubit, gate, and measurement during the syndrome extraction circuit. 

QLDPC codes are promising candidates for the implementation of QEC~\cite{breuckmann2021quantum}. The low-density parity-check (LDPC) constraint states that for a given family of codes, the number of syndrome (data) qubits connected to each data (syndrome) qubit is bound by a constant independent of the code size. This allows the depth of the syndrome extraction circuit to be constant as the code size scales up, which is crucial for realistic implementation. 

The most widely studied example is the surface code~\cite{raussendorf2007fault, fowler2009high, fowler2012surface}, where each data (syndrome) qubit is connected to its nearest-neighbor syndrome (data) qubits in a 2-dimensional lattice. Surface codes have several advantages, such as a high circuit-level error threshold and a planar layout of qubits, leading to the recent experimental realization of QEC below the threshold~\cite{google2025quantum}. The drawback of surface codes is that they have asymptotically vanishing rates: as the lattice size $L\rightarrow \infty$, both the encoding rate $k/n = \Theta(L^{-2})$ and the distance rate $d/n = \Theta(L^{-1})$ vanish. This saturates the fundamental bounds on the parameters of QEC codes with 2-dimensional local connectivity~\cite{bravyi2010tradeoffs}.

By removing the constraint of local connectivity, various families of QLDPC codes with high encoding rate $k/n$ and distance rate $d/n$ are found~\cite{tillich2013quantum, panteleev2021degenerate, panteleev2021quantum, breuckmann2021balanced, leverrier2022quantum}. The QEC performance of these codes depends on the choice of syndrome extraction circuit and the decoder implementation, which are more complicated than those for surface codes. The goal of this paper is to provide a modular framework for constructing the circuits and evaluating the decoders for the QLDPC codes with non-local connectivity.

\section{QLDPC codes} \label{sec:qeccodes}

\begin{figure*}[ht!]
  \centering
  \includegraphics[width=.9\textwidth]{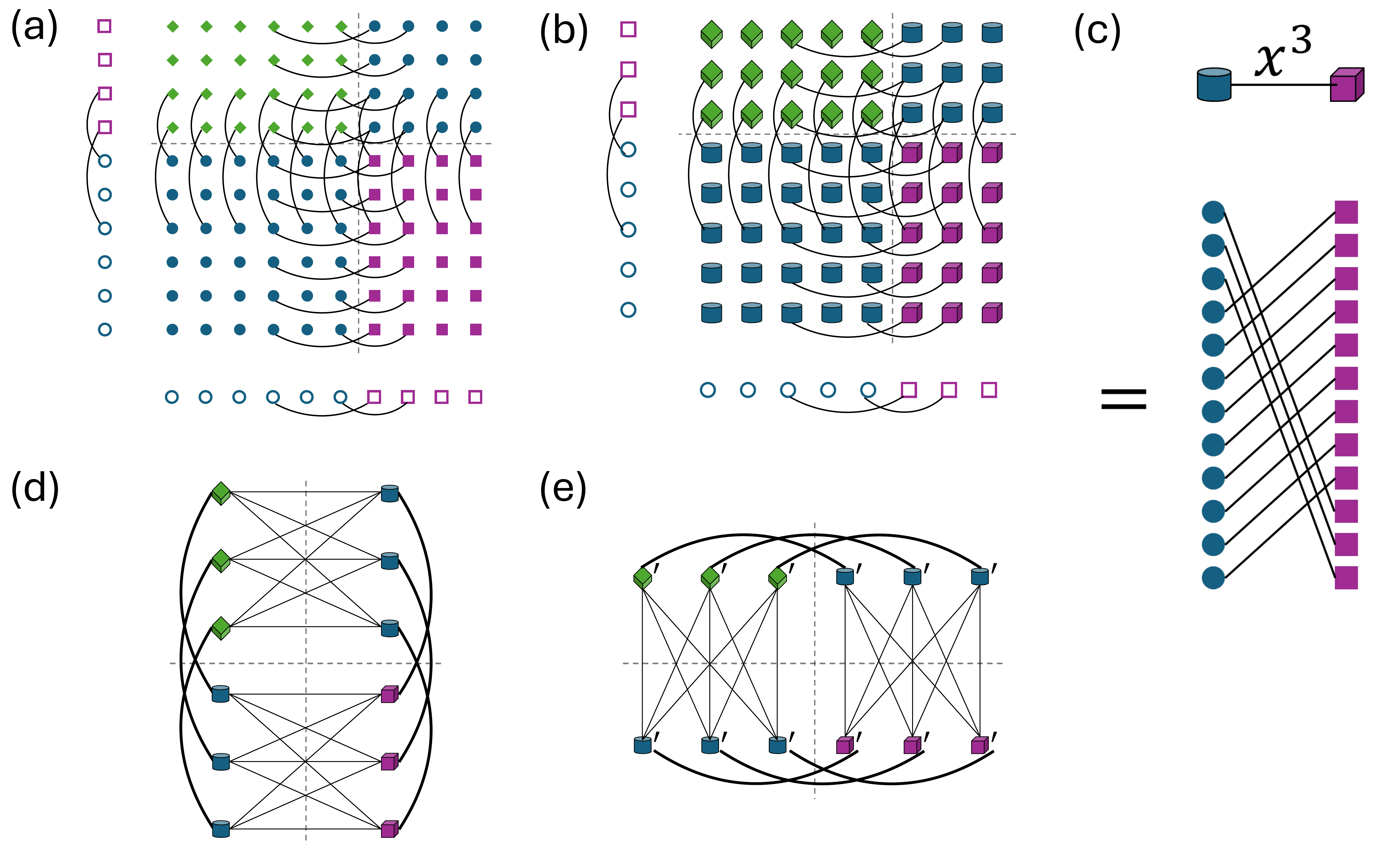}
  \caption{\textbf{QLDPC codes}. Filled blue circles, green diamonds, and purple squares represent the data, $Z$-type syndrome, and $X$-type syndrome qubits, respectively. 3-dimensional symbols represent the lifted nodes, each of which contains $l$ qubits of the corresponding type, where $l$ is the lift size. \textbf{(a)} Tanner graph of a hypergraph product code, constructed from a product of two classical codes. Empty blue circles and purple squares represent the data and check bits of the classical codes, respectively. \textbf{(b)} Tanner graph of a quasi-cyclic lifted product code. Each edge is assigned a monomial $x^s$ ($s\in[l]$, $x^l=1$). \textbf{(c)} Example of unraveling the edge between two lifted nodes ($l=12$). \textbf{(d)(e)} Tanner graph of a balanced product cyclic code. Each edge is assigned a polynomial with terms $x^s$. In (d), vertical edges have a one-to-one mapping. In (e), the qubits are ``shuffled'' within each quadrant such that horizontal edges have a one-to-one mapping.}
  \label{fig:qldpccodes}
\end{figure*}

We focus on QLDPC codes constructed from a product of two classical codes. The product structure admits a unified method for constructing the syndrome extraction circuit, as will be explained in Sec.~\ref{sec:syndcircuits}.

\subsection{Hypergraph product (HGP) codes}

HGP codes~\cite{tillich2013quantum} are the most straightforward example of QLDPC codes with product structure. Consider two classical LDPC codes with parity check matrix $H_\alpha$ and parameters $[n_\alpha,\:k_\alpha,\:d_\alpha]$ ($\alpha=1,2$); we also define $r_\alpha:=n_\alpha-k_\alpha$. The classical check (data) bits are labeled as $c_\alpha^{i_\alpha}$ ($v_\alpha^{j_\alpha}$), where $i_\alpha \in [r_\alpha]$ ($j_\alpha\in[n_\alpha]$), such that $H_\alpha^{i_\alpha j_\alpha}=1$ if the $i_\alpha$th check $c_\alpha^{i_\alpha}$ has support on the $j_\alpha$th data bit $v_\alpha^{j_\alpha}$ and $H_\alpha^{i_\alpha j_\alpha}=0$ otherwise. The product of these two classical codes yields the HGP code as illustrated in Fig.~\ref{fig:qldpccodes}(a), where $H_1$ and $H_2$ represent the horizontal and vertical edges of the Tanner graph, respectively. The data qubits are labeled as $v_1^{j_1} v_2^{j_2}$ (lower-left quadrant with respect to the gray dashed lines) and $c_1^{i_1} c_2^{i_2}$ (upper-right quadrant), the $Z$-type syndrome qubits are labeled as $v_1^{j_1} c_2^{i_2}$ (upper-left quadrant), and the $X$-type syndrome qubits are labeled as $c_1^{i_1} v_2^{j_2}$ (lower-right quadrant). The HGP code has the parity check matrices
\begin{equation}
\begin{aligned}
     H_Z &= \begin{pmatrix}
        H_2 \otimes I_{n_1} \:|\: I_{r_2} \otimes H_1^T
    \end{pmatrix}, \\
    H_X &= \begin{pmatrix}
        I_{n_2} \otimes H_1 \:|\: H_2^T \otimes I_{r_1}
    \end{pmatrix},
\end{aligned}
\end{equation}
where $I_n$ denotes the $n$ by $n$ identity matrix, such that each $Z$-type check corresponding to $v_1^{j_1}c_2^{i_2}$ is supported by data qubit $v_1^{j_1} v_2^{j_2}$ ($c_1^{i_1} c_2^{i_2}$) if and only if $H_2^{i_2 j_2}=1$ ($H_1^{i_1 j_1}=1$), and similar holds for $X$-type checks. The code parameters are given by
\begin{equation*}
    [[n=n_1n_2+r_1r_2, \: k=k_1k_2,\: d=\text{min}(d_1,d_2)]].
\end{equation*}
As a family of classical LDPC codes typically has constant encoding rate and distance rate, i.e., $k_\alpha/n_\alpha=\Theta(1)$ and $d_\alpha/n_\alpha=\Theta(1)$, the rates of the family of corresponding HGP codes scale as $k/n=\Theta(1)$ and $d/n=\Theta(1/\sqrt{n})$. 

Following the convention from the literature~\cite{grospellier2021combining, tremblay2022constant, xu2024constant,huang2024increasing}, we use the classical LDPC codes where each data bit is supported by 3 checks and each check supports 4 data bits. This yields a family of HGP codes with a fixed encoding rate $k/n=1/25$, where each data (syndrome) qubit is connected to 6 to 8 syndrome (data) qubits. The HGP codes are known to perform well when the underlying classical codes have a large girth, i.e., the length of the shortest cycle in the Tanner graph~\cite{grospellier2021combining, raveendran2022finite}. We implement the method in Ref.~\cite{grospellier2019constant} to optimize the distance and girth of classical LDPC codes in the \verb|generate_ldpc| function (feature of \texttt{QUITS} outside the shaded area in Fig.~\ref{fig:QUITS}).

\subsection{Quasi-cyclic lifted product (QLP) codes}

Lifted product codes are constructed by combining the hypergraph product with the lifting procedure~\cite{panteleev2021degenerate, panteleev2021quantum}. They encode a larger number of logical qubits with a larger code distance compared to HGP codes with a similar number of data qubits, as the lifting procedure reduces the symmetries of the product structure~\cite{breuckmann2021quantum}. 

Here we consider the QLP codes~\cite{panteleev2021quantum,raveendran2022finite}, which are obtained by taking the product of two base matrices defined over the quotient polynomial ring $\mathbb{R}[x]/(x^l-1)$ and then ``lifting'' each monomial entry into a $l$ by $l$ circulant matrix, where $l$ is the lift size~\cite{xu2024constant}. Specifically, given the base matrices $B_\alpha$ of size $m_\alpha$ by $n_\alpha$ ($\alpha=1,2$), the parity check matrices of the QLP code are given by
\begin{equation}
\begin{aligned}
     H_Z &= \begin{pmatrix}
        B_2 \otimes I_{n_1} \:|\: I_{m_2} \otimes B_1^T
    \end{pmatrix}, \\
    H_X &= \begin{pmatrix}
        I_{n_2} \otimes B_1 \:|\: B_2^T \otimes I_{m_1}
    \end{pmatrix}.
\end{aligned}
\end{equation}
Then, each monomial entry $x^s$ ($s \in[l]$, $x^l=1$) is replaced by $l$ by $l$ matrix $M$ with entries $M^{ij}=\delta_{i,(j+s)\:\text{mod}\:l}$, where $\delta$ is the Kronecker delta symbol, and each zero entry is replaced by $l$ by $l$ matrix with all-zero entries~\cite{xu2024constant}.  Note that if $B_\alpha^{i_\alpha j_\alpha} = x^s$, then $(B_\alpha^{T})^{j_\alpha i_\alpha} = x^{(l-s)}$. The number of data qubits is given by $n=l(n_1n_2+m_1m_2)$ and the number of logical qubits is upper bounded as $k \leq n-l(n_1m_2+n_2m_1)$. 

The Tanner graph of a QLP code can also be understood from its product structure~\cite{xu2024constant}, as shown in Fig.~\ref{fig:qldpccodes}(b). The data qubits are assigned labels $(v_1^{j_1} v_2^{j_2}, s)$ (lower-left quadrant) and $(c_1^{i_1} c_2^{i_2}, s)$ (upper-right quadrant), the Z-type syndrome qubits are assigned $(v_1^{j_1} c_2^{i_2}, s)$ (upper-left quadrant), and the X-type syndrome qubits are assigned $(c_1^{i_1} v_2^{j_2}, s)$ (lower-right quadrant), where $i_\alpha\in[m_\alpha]$, $j_\alpha\in[n_\alpha]$, and $s\in[l]$. Each $Z$-type check corresponding to $(v_1^{j_1} c_2^{i_2}, s_z)$ is supported by data qubit $(v_1^{j_1} v_2^{j_2}, s)$ [$(c_1^{i_1} c_2^{i_2}, s)$] if and only if $B_2^{i_2 j_2} = x^{(s_z-s) \: \text{mod}\:l}$ [$(B_1^T)^{i_1 j_1} = x^{(s-s_z) \: \text{mod}\:l}$], and similar holds for $X$-type checks. An example of unraveling the Tanner graph edge between two ``lifted'' nodes is visualized in Fig.~\ref{fig:qldpccodes}(c). 

The family of QLP codes used in this work is defined from the 3 by 5 base matrices in Eqs.~(5) and (6) of Ref.~\cite{xu2024constant}, such that each data (syndrome) qubit is connected to 6 to 10 syndrome (data) qubits. Other base matrices can be straightforwardly incorporated. 

\begin{figure*}[ht!]
  \centering
  \includegraphics[width=\textwidth]{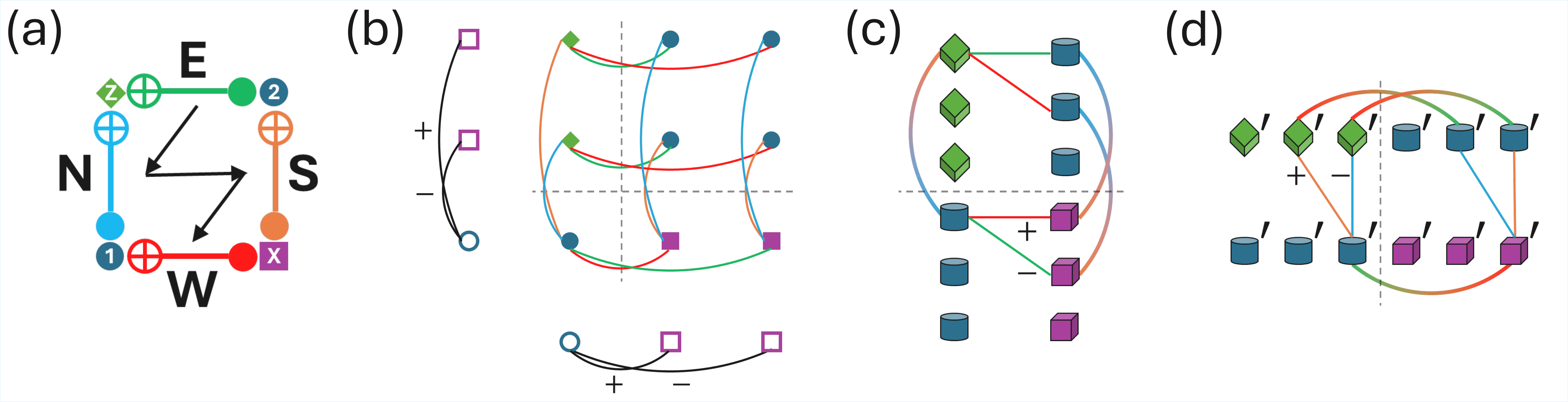}
  \caption{\textbf{Syndrome extraction circuit.} \textbf{(a)} A \textit{qubit 4-cycle}, or a cycle of (data qubit)-(Z syndrome qubit)-(data qubit)-(X syndrome qubit), should consist of edges of all 4 directions. We use the convention in Ref.~\cite{tremblay2022constant} and set the ordering of the directions as E$\rightarrow$N$\rightarrow$S$\rightarrow$W. \textbf{(b)} Assignment of the signs of the classical Tanner graph's edges, which determines the directions of the quantum Tanner graph's edges. Note that the colors represent the directions as matched in (a) and are irrelevant to the edge coloration. \textbf{(c)}\textbf{(d)} Directions of the edges in a BPC code's Tanner graph. Horizontal edges are assigned directions E and W in (c), and vertical edges are assigned directions N and S in (d).}
  \label{fig:circuit}
\end{figure*}

\subsection{Balanced product cyclic (BPC) codes}

Balanced product codes are another example of codes constructed by reducing the symmetries of the hypergraph product~\cite{breuckmann2021balanced}. Given two classical codes with a common symmetry group, the balanced product factors out the action of the group on their hypergraph product~\cite{breuckmann2021quantum}.

This work considers the family of BPC codes introduced in Ref.~\cite{tiew2024low}, although other BPC codes can be straightforwardly implemented. The codes are obtained from the balanced product $\otimes_H$ of two copies of a cyclic group $C_{3q}=\langle x \rangle$, where a smaller cyclic subgroup $H=C_q=\langle x^3 \rangle$ is factored out. The size of cyclic group $l:=3q$ can be considered as the lift size. Given two polynomials $p_1(x)$ and $p_2(x)$, the parity check matrices of the BPC code are given by
\begin{equation}\label{eq:bpcH}
\begin{aligned}
    H_Z &= \begin{pmatrix}
        e \otimes_H p_2(x) \: | \: (p_1(x) \otimes_H e)^T 
    \end{pmatrix},\\
    H_X &= \begin{pmatrix}
        p_1(x) \otimes_H e \: | \: (e \otimes_H p_2(x))^T
    \end{pmatrix},
\end{aligned}
\end{equation}
where $e$ is the identity element. The balanced product $\otimes_H$ of two polynomials is a 3 by 3 matrix with polynomial entries. The specific rules of $\otimes_H$ can be found in Ref.~\cite{tiew2024low}; crucially, $p_1(x) \otimes_H e$ only involves terms $x^{s}$ where $s\equiv 0 \:\text{mod}\:3$, and $e \otimes_H p_2(x)$ is simply a diagonal matrix with $p_2(x)$ at all entries. Note that transpose works analogously to QLP codes. Then, each term $x^s$ ($s \in[l]$, $x^l=1$) is lifted to $l$ by $l$ matrix $M$ with entries $M^{ij}=\delta_{i,(j+s)\:\text{mod}\:l}$, and each zero entry is lifted to $l$ by $l$ matrix with all-zero entries. This construction yields a family of codes with parameters~\cite{tiew2024low}
\begin{equation*}
    [[n=18q, \: k=8, \: d\leq 2q]].
\end{equation*}
Although the encoding rate $k/n$ approaches zero asymptotically, the distance rate $d/n$ is relatively high at least for known examples~\cite{tiew2024low}. Also, if each of $p_1(x)$ and $p_2(x)$ has 3 terms, each data (syndrome) qubit is connected to only 6 syndrome (data) qubits, making this family of codes a promising candidate for implementation.

The Tanner graph of the BPC codes is visualized in Fig.~\ref{fig:qldpccodes}(d) and (e). In Fig.~\ref{fig:qldpccodes}(d), the qubits are arranged such that each node (containing $l$ qubits) represents the row or column of the parity check matrices of Eq.~\eqref{eq:bpcH}, which are 3 by 6 matrices with polynomial entries. Specifically, the $j$th node of the lower-left [upper-right] quadrant is assigned the $j$th [$(j+3)$th] column that represents the data qubits $(j=1,2,3)$, and the $i$th node of the upper-left (lower-right) quadrant is assigned the $i$th row of $H_Z$ ($H_X$) that represents the $Z$($X$)-type syndrome qubits ($i=1,2,3$). As $e \otimes_H p_2(x)$ is diagonal, the \textit{vertical} edges enjoy a one-to-one mapping between the nodes; i.e., each $Z$($X$)-type check corresponding to a qubit in the $i$-th node of the upper-left (lower-right) quadrant is supported by data qubits in the $i$-th node of the lower-left (upper-right) quadrant. 

The $3l$ qubits of the 3 nodes in each quadrant can be ``shuffled'' to yield the Tanner graph in Fig.~\ref{fig:qldpccodes}(e), where the \textit{horizontal} edges have a one-to-one mapping. Specifically, the $s$th qubit of the $i$th node ($s\in [l]$, $i\in [3]$) is shuffled to the $s'$th qubit of the $i'$th node, where $s=3(m-1)+r$ ($m\in[q]$, $r\in[3]$), $s'=q(i-1)+m$, and $i'=r$. This allocation of qubits guarantees that the $Z$($X$)-type check corresponding to a qubit in the $i'$-th node of the upper-left (lower-right) quadrant is supported by data qubits in the $i'$-th node of the upper-right (lower-left) quadrant, as \mbox{$p_1(x)\otimes_H e$} matrix only contains terms of which power is multiple of 3. As will be shown in Sec.~\ref{sec:syndcircuits}, the allocation of qubits that allows one-to-one mapping for edges in the horizontal or vertical direction is crucial to the design of the syndrome extraction circuit. 

The family of BPC codes used in this work is defined from the weight 3 polynomials in Table 1 of Ref.~\cite{tiew2024low}, such that each data (syndrome) qubit is connected to 6 syndrome (data) qubits. Codes constructed from polynomials of larger weight or other cyclic groups can be straightforwardly implemented.

\section{Syndrome extraction circuits}
\label{sec:syndcircuits}

The syndrome extraction circuits of the QLDPC codes in Sec.~\ref{sec:qeccodes} can be constructed in a unified way. We use a variant of the cardinal circuit developed in Ref.~\cite{tremblay2022constant}, which performed the first circuit-level simulations of HGP codes. Our method can be applied to QLP codes~\cite{xu2024constant} and BPC codes, while also reducing the depth of the syndrome extraction circuit of small HGP codes (see Appx.~\ref{app:depth}). The syndrome extraction depths of all QLDPC codes considered in this work are shown in Appx.~\ref{app:depth}.

A syndrome extraction circuit, which is performed at each QEC cycle, consists of the following steps: (i) prepare the $Z(X)$-type syndrome qubits at the $\ket{0}$ ($\ket{+}$) state, (ii) apply the layers of CNOT gates sequentially, and (iii) measure the syndrome qubits in the respective basis. In each entangling gate layer, a qubit can have at most one gate applied to it. For optimal QEC performance, the application of CNOT gates needs to be parallelized as much as possible, reducing the number of layers.

The entangling gates are parallelized in the cardinal circuit of the HGP codes as follows~\cite{tremblay2022constant}. First, a direction (E, N, S, W) is assigned to each edge of the Tanner graph, where E or W (N or S) is assigned to a horizontal (vertical) edge. Each cycle of (data qubit)-($Z$ syndrome qubit)-(data qubit)-($X$ syndrome qubit), hereafter referred to as \textit{qubit 4-cycle}, should consist of edges of all directions, as shown in Fig.~\ref{fig:circuit}(a). Then, for each direction, we apply \textit{edge coloration} to the subgraph of the Tanner graph that consists of edges of the given direction, such that no two edges that share a vertex have the same color. The CNOT gates corresponding to edges with the same direction and color are applied simultaneously. Thus, the total number of entangling gate layers, or \textit{syndrome extraction depth}, is given by the sum of the number of colors for each direction. The ordering of directions is crucial for extracting syndromes correctly. An improper ordering can entangle syndrome qubits and lead to random syndrome outcomes (see Appx.~B of Ref.~\cite{fowler2012surface}). Here, we follow Ref.~\cite{tremblay2022constant} and determine the ordering as $\text{E} \rightarrow \text{N} \rightarrow \text{S} \rightarrow \text{W}$; i.e., all gates of direction E are applied, then N, S, and W. 

Now we explain how the horizontal edges are assigned directions E or W, such that each qubit 4-cycle consists of both E and W edges. Vertical edges are analogously assigned directions N or S; see Fig.~\ref{fig:circuit}(b) for visualization. For HGP codes, data qubits $v_1^{j_1}v_2^{j_2}$, Z-type syndrome qubits $v_1^{j_1}c_2^{i_2}$, data qubits $c_1^{i_1}c_2^{i_2}$, and X-type syndrome qubits $c_1^{i_1}v_2^{j_2}$ are allocated in the lower-left, upper-left, upper-right, and lower-right quadrants, respectively. Thus, for each edge between nodes $v_1^{j_1}$ and $c_1^{i_1}$ of the underlying classical code's Tanner graph, the horizontal edges between qubits $v_1^{j_1}v_2^{j_2}$ and $c_1^{i_1}v_2^{j_2}$ in the lower two quadrants and the horizontal edges between qubits $v_1^{j_1}c_2^{i_2}$ and $c_1^{i_1}c_2^{i_2}$ in the upper two quadrants should have opposite directions, for all connected pairs of $v_2^{j_2}$ and $c_2^{i_2}$. To enforce this restriction, we assign a sign ($-$ or $+$) to each edge in the Tanner graph of the underlying classical code. Then, for each $-$ ($+$) edge, we assign direction E (W) to the corresponding horizontal edges of the quantum code's Tanner graph between the lower two quadrants. The edges between the upper two quadrants are automatically assigned the opposite direction.

To minimize the syndrome extraction depth, a balance between the directions is desired~\cite{tremblay2022constant}; for each vertex, the horizontal (vertical) incident edges need to be distributed to E and W (N and S) directions as evenly as possible. We use a heuristic algorithm introduced in Appx.~\ref{app:depth} to balance the signs of the classical Tanner graph's edges incident to each vertex. A key difference of our method from Ref.~\cite{tremblay2022constant} is that the assignment of signs does not involve the positions of the vertices (after permutation). This flexibility allows for finding balanced solutions more easily, leading to a reduction in the syndrome extraction depth of $[[225, 9, 6]]$ HGP code from 12 to 8. We note that Ref.~\cite{tremblay2022constant} defines the balanced ordering of the classical Tanner graph's vertices; however, an explicit method for finding such an ordering is not provided. 

For QLP and BPC codes, which also have a product structure, the qubits are similarly allocated to four quadrants, as illustrated in Fig.~\ref{fig:qldpccodes}. Thus, the same procedure as described above assigns the direction to each edge between two lifted nodes, each of which represents a set of $l$ qubits. The edges between the two sets of $l$ qubits in the quantum code's Tanner graph [see Fig.~\ref{fig:qldpccodes}(c)] are assigned the same direction as the corresponding edge between the lifted nodes.

The balanced product structure in BPC codes requires additional care in assigning the directions, as shown in Fig.~\ref{fig:circuit}(c) and (d). In (c), the horizontal edges below the dashed line can be thought of as the edges of a ``classical'' Tanner graph, to which signs are assigned. Due to the one-to-one mapping of the vertical edges, the pair of horizontal edges is unambiguously defined for each qubit 4-cycle. Thus, E and W directions can be assigned using the same method. In (d), the vertical edges between the ``shuffled'' nodes are assigned directions analogously, which is again enabled by the one-to-one mapping of the horizontal edges. This completes the assignment of directions such that each qubit 4-cycle has edges of all 4 directions.

\section{Decoder implementation}\label{sec:decoder} 

\begin{figure*}[ht!]
  \centering
  \includegraphics[width=\textwidth]{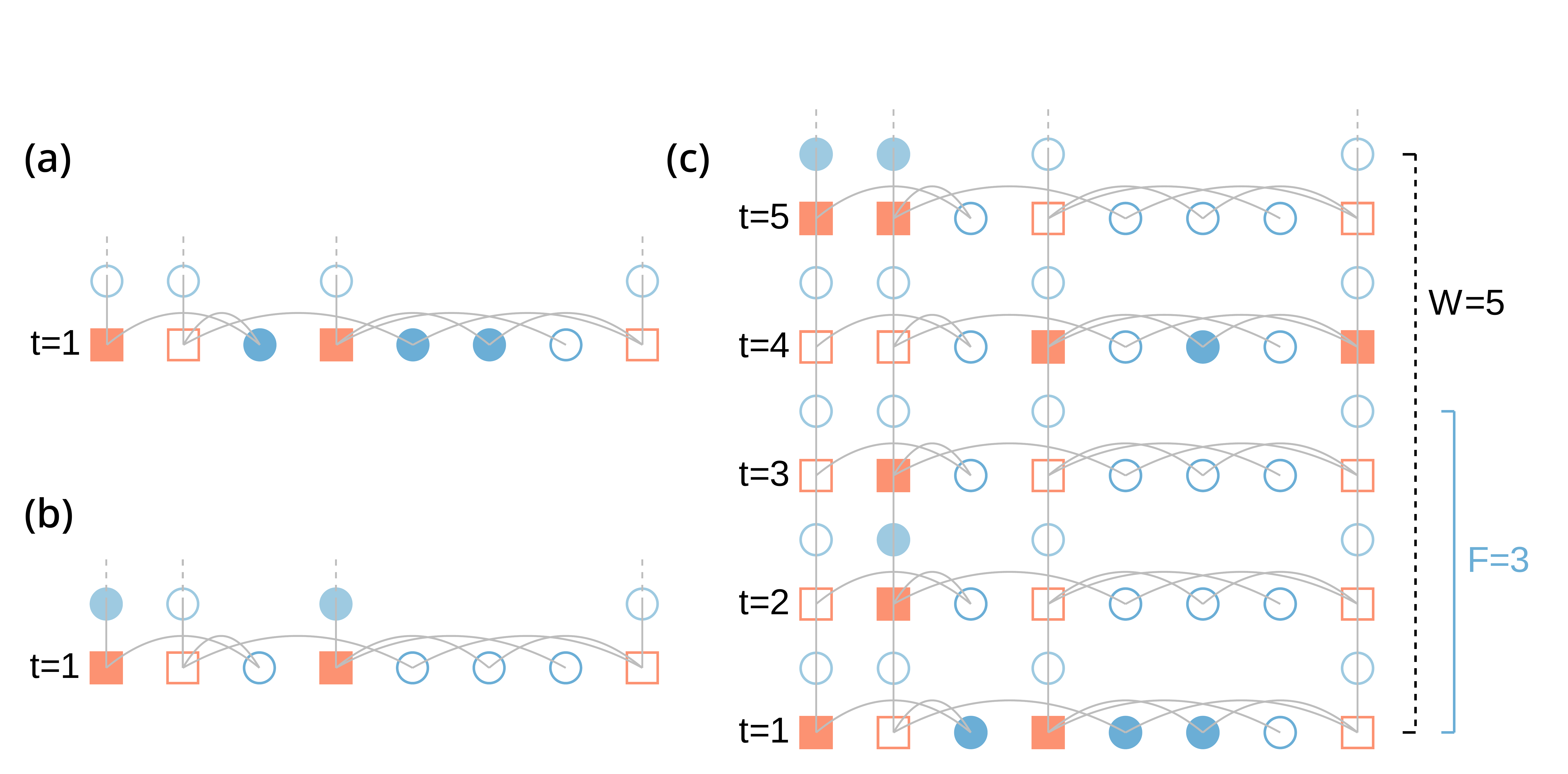}
  \caption{\textbf{Space-time decoding graph.} Tanner graph constructed from a detector error matrix. Each square represents a detector. Each circle represents a possible error mechanism. Filled circles are errors that occur. Filled squares are the detectors flipped by the errors. For simplicity, we show a phenomenological noise model where there are only data qubit errors (blue circles) and measurement errors (light blue circles on the vertical edges). Dashed gray lines mark the time-like boundary of the decoding graph, which connects error mechanisms to detectors of future time steps they support. \textbf{(a)(b)} Two possible combinations of errors that create the same detector flips. If single-shot error correction is performed and the error shown in (a) occurs, a minimum-weight decoder will apply the correction shown in (b) as it is the error combination with the lowest weight that produces the given flipped detector set. \textbf{(c)} A decoding window for a $(5,3)$ sliding-window decoding. For each decoding window, we pass the detector flips of $W$ consecutive time steps to an inner decoder. Based on the correction suggested by the inner decoder, we apply the correction that flips the detectors in the first $F$ time steps. By increasing the size of the decoding window, more detector information is included in the decoding window. Thus, the decoder can correct the error shown in (a) that occurs at the beginning of the decoding window. }
\label{fig:sliding-window}
\end{figure*}

In this section, we introduce a general sliding-window decoding framework~\cite{huang2024increasing}. For decoding under a circuit-level noise model, we enable implementing various inner decoders within the sliding-window decoding framework, as illustrated in Fig.~\ref{fig:QUITS}. The modular structure of \QUITS\ that decouples the inner decoder from the sliding-window decoding framework facilitates the development and testing of various custom-defined inner decoders.

\subsection{Detectors}

Before we discuss sliding-window decoding, we first introduce detectors. Detectors are products of measurements that return $+1$ in the absence of errors, and $-1$ when errors are detected. As each detector can only be flipped by certain combinations of physical errors, a decoder can locate and correct errors based on the detector values. We construct $P$-type \textit{detectors} $D^{i}_{P}(t)$ at round $t$ by taking the parity of two consecutive measurements $M^{i}_{P}(t)$ at round $t$ and $M^{i}_{P}(t+1)$ at round $t+1$ of the $i$th $P$-type check, i.e., 
\begin{equation}
    D^{i}_{P}(t) := M^{i}_{P}(t) M^{i}_{P}(t+1),
\end{equation}
where $i$ denotes the index of the detector. 
For the initial round $t=0$, we define the detectors to be $D^{i}_{P}(t=0) := M_P^i(t=1)$. 

Similarly to parity check matrices, we construct \textit{detector error matrices}. In a $P$-type detector error matrix $\widetilde{H}_P$ for the whole circuit, each column represents a possible error mechanism that flips $P$-type detector, while each row represents a $P$-type detector. Each error mechanism is an incident of error, such as qubit error, measurement error, and gate error, that can occur independently. Two error mechanisms are equivalent if they generate the same syndromes and logical observables. A matrix element $\widetilde{H}_P^{ij}=1$ if the $i$th detector is flipped by the $j$th error mechanism. The detector error matrices can be readily converted from the detector error model in Stim \cite{gidney2021stim,gong2024toward,higgott2023improved}. When constructing the detector error matrix, equivalent error mechanisms are merged into a single error mechanism with an updated error rate. An example Tanner graph constructed from a detector error matrix is shown in Fig.~\ref{fig:sliding-window}(c). 

There are two categories of error mechanisms captured by detector error matrices: (i) space-like errors, which only flip the detectors of the same round (e.g. the blue circles in Fig.~\ref{fig:sliding-window}(c)), and (ii) time-like errors, which flip the detectors in two consecutive rounds (e.g. the light blue circles on the vertical edges in Fig.~\ref{fig:sliding-window}(c)). In a phenomenological noise model, which includes only data qubit errors and measurement errors, space-like errors are the data qubit errors, and time-like errors are the measurement errors. In a circuit-level noise model, where errors may occur during the execution of the syndrome extraction circuit, idling errors on data qubits can also cause time-like errors. For example, if an idling $X$-type error on a data qubit occurs between the CNOT gates in the syndrome extraction circuit at round $t$, only a subset of the $Z$-type checks supported by this data qubit will have their signs flipped at round $t$, and the remaining checks' signs will be flipped at round $t+1$, together causing a time-like error. Similarly, gate errors can cause both space-like and time-like errors, whereas measurement errors can only cause time-like errors. To achieve accurate decoding, using a decoding graph that includes both space-like and time-like errors is necessary, as described in Fig.~\ref{fig:sliding-window}. Note that a single error mechanism cannot flip detectors from more than two rounds.

\subsection{Sliding-window decoding}

Due to the existence of time-like errors, multiple rounds of syndrome measurement results need to be decoded together, which necessitates sliding-window decoding. While some QLDPC codes support single-shot error correction, where error correction is performed immediately after each round of syndrome extraction \cite{campbell2019theory, fawzi2020constant, dennis2002topological}, sliding-window decoding has been shown to outperform single-shot error correction for HGP and QLP codes even under a phenomenological noise model~\cite{huang2024increasing}. The performance degradation of single-shot error correction arises from the decoder's inability to distinguish between space-like errors and time-like errors. For example, Fig.~\ref{fig:sliding-window}(a) and (b) present two example error combinations that generate the same syndrome in a phenomenological noise model. Given only one round of syndrome extraction, if independent measurement errors and data qubit errors occur at the same probability, a minimum-weight decoder always chooses to apply the error combination in (b) as it contains fewer independent errors. 

Sliding-window decoding is capable of distinguishing space-like and time-like errors. In general, sliding-window decoding is implemented by dividing syndromes from different rounds into overlapping decoding windows that are individually passed to an inner decoder. Only the correction for the portion of the window that does not overlap with the next is applied, while the syndromes in the next decoding window are updated accordingly~\cite{huang2024increasing}. An example of a decoding window under a phenomenological noise model with $W=5$ rounds of syndrome extraction is shown in Fig.~\ref{fig:sliding-window}(c). In this example, errors at the beginning of the window can be reliably corrected. Note that data-qubit errors occurring at the end of the window cannot be distinguished from measurement errors, for the same reason as the case in Fig.~\ref{fig:sliding-window}(a) and (b). To ensure reliability, only the correction for the first $F$ rounds of syndrome is applied, while the remaining syndrome corrections are deferred to the next decoding window. Fig.~\ref{fig:sliding-window}(c) shows an example of $F=3$.

Formally, we parametrize space-time sliding-window decoding by $(W,F)$, where $W$ denotes the number of detector rounds included in each window and $F$ denotes the number of rounds in correction that we commit~\cite{huang2024increasing}. To decode the $w$th window, we pass into the inner decoder a submatrix of the $P$-type detector error matrix consisting of rows labeled from $(w-1)Fc$ to $((w-1)F+W-1)c$ and columns that have non-trivial support on these rows, where $c$ denotes the number of $P$-type checks we measure per round and $w=1,\cdots,N$ with $N$ as the total number of decoding windows. The inner decoder produces a correction based on these submatrices and the corresponding error rates. Before the last window, from the correction, the error mechanisms affecting $D_{P}^{i}(t)$ where $i=1,\cdots,c$ and $t=(w-1)F,\cdots,wF-1$ are committed. The following set of detectors $\{D^{i}_{P}(wF)\}$ is updated accordingly. In the last window, all the correction is applied. See also Refs.~\cite{gong2024toward, scruby2024high, berent2024analog, lin2025single} where a similar decoding scheme is applied to other QLDPC codes.

Before discussing inner decoders, we highlight that \QUITS\ features a modular structure that separates inner decoders from the space-time sliding-window decoding framework, unlike previous implementations where the inner decoder is tightly integrated into the framework \cite{gong2024toward}. This separation enables convenient testing of different inner decoder algorithms.

\subsection{Inner Decoder}
While the outer sliding-window decoder characterizes the relationship between possible error mechanisms and the detectors, we must still decode the detector error matrix within a window. One common class of decoders that can be applied to infer an error pattern is the belief-propagation(BP)-based decoders \cite{panteleev2021degenerate,hillmann2024localized}.

\subsubsection{Belief Propagation Decoding}
The BP decoder is an iterative message-passing algorithm widely used in classical error correction, particularly for LDPC codes. When applied to the Tanner graph representing the detector error matrix, BP iteratively exchanges messages between the detectors and the error mechanisms, attempting to identify a probable error pattern that matches the observed detector flips. If BP successfully finds such an error pattern, it is said to \textit{converge}. The computational complexity of BP decoding is $\mathcal{O}(\mathcal{I}n)$, where $\mathcal{I}$ denotes the number of message passing iterations.

When applied to quantum codes, BP decoding faces several known challenges that degrade its performance. In particular, short cycles in the factor graph can hinder BP convergence. Quantum LDPC codes inherently contain many loops due to degeneracy, making the factor graph far from locally tree-like and obstructing the convergence of BP. Under circuit-level noise, this issue is further amplified by correlated errors introduced around quantum gates. A single physical error can propagate through two-qubit gates and trigger multiple syndrome measurements, effectively creating additional short cycles in the Tanner graph of the detector error model.
Another challenge is the presence of syndrome measurement errors, where a single measurement error can activate two consecutive detectors \cite{gong2024toward}. 

\subsubsection{Heuristics for improving BP}
Motivated by issues of BP on quantum Tanner graphs, some post-processors of BP have been proposed. In practice, these BP-based post-processing decoders substantially improve reliability in the presence of correlated fault patterns, which is crucial for decoding QLDPC codes under realistic circuit-level noise. In this section, we briefly review some of these post-processing schemes.

In cases where BP does not converge, it still produces soft estimates on the likelihoods of all the error mechanisms. In this case, order-statistics decoding (OSD) can be applied as a post-processing step to improve the BP decoding performance. This approach, known as BP with order-statistics decoding (BP-OSD), takes the BP soft output and computes a probable error pattern that matches the detector flips by solving a system of linear equations. Even a low-order OSD post-processing yields significant gains. BP-OSD with order zero has a computational complexity of $\mathcal{O}(n_e^3)$, where $n_e$ denotes the number of error mechanisms. For a more comprehensive introduction to BP-OSD, we refer the reader to Ref.~\cite{panteleev2021degenerate}.

Another decoder supported by \QUITS{}, imported from \cite{Roffe_LDPC_Python_tools_2022}, is belief propagation with localized statistics decoding (BP-LSD) \cite{hillmann2024localized}, a more recent decoding scheme that improves upon BP-OSD.
To improve the efficiency of OSD, BP-LSD identifies small, isolated clusters consisting of check nodes and fault nodes in the Tanner graph and then decodes them separately using OSD in parallel.
Clusters are grown starting from the activated detector nodes, with their growth guided by BP. They are merged when necessary and expanded until a local solution can be identified.
BP-LSD has been shown to achieve the same level of accuracy as BP-OSD while enabling parallel decoding of the cluster solutions.

In \QUITS, we provide built-in support of the BP-OSD and BP-LSD decoders from the LDPC package \cite{Roffe_LDPC_Python_tools_2022} within the space-time sliding-window framework, alongside a general decoding framework that accommodates any customized decoders.

\section{Results} \label{sec:results}

In this section, we simulate the circuit-level performance of the memory of various QLDPC codes. The HGP codes used in this section are constructed from the product of classical codes, each associated with a random $(3,4)$-regular Tanner graph. These classical codes are obtained from the procedure described in Ref.~\cite{grospellier2021combining}, implemented in the \texttt{generate\_ldpc} function of \QUITS. By sampling the underlying classical codes with a minimum girth 6, we choose HGP codes with the largest code distance. We denote these codes by $n$ as HGP225, HGP625, and HGP900. With \QUITS\ circuit generator, we found circuits with minimum syndrome extraction depth $\min{d_c}=8$ for HGP225 and $12$ for HGP625 and HGP900. The QLP and BPC codes are brought from Refs.~\cite{xu2024constant} and \cite{tiew2024low}, respectively, as explained in Sec.~\ref{sec:qeccodes}. The minimum syndrome extraction depth found using \QUITS\ was $d_c=12$ and $8$ for the QLP and BPC code family, respectively. See Appx.~\ref{app:depth} for a detailed discussion of the syndrome extraction depth $d_c$. 

For the noise model, we use the standard circuit-level depolarizing noise model characterized by the physical error rate $p$. At each time step, all qubits experience a single-qubit depolarizing channel where all single-qubit Pauli errors $X$, $Y$, and $Z$ occur at an equal probability $p/3$ when idling or subjected to single-qubit gates. All the CNOT gates induce two-qubit depolarizing channels on the involved qubits, where the $15$ possible two-qubit Pauli errors from the set $\{I,X,Y,Z\}^{\otimes2}/I\otimes I$ occur at an equal probability $p/15$. Reset and measurement errors on syndrome qubits for $Z$($X$)-type error detection are modeled by single qubit bit-flip (phase-flip) channels where $X$($Z$) errors occur at a probability $p$. Each of these single-qubit or two-qubit Pauli errors at different locations represents an error mechanism. Independent errors on syndrome qubits may propagate through CNOT gates, leading to correlated errors on multiple data qubits. These correlated errors, known as hook errors, can significantly degrade the code performance. \cite{dennis2002topological, tomita2014low}. 

For each code, we perform $T=16$ rounds of noisy syndrome extraction circuits. 
At the end of the circuit, each physical qubit undergoes a noiseless measurement, with the results combined into the final stabilizer and logical operator measurements. We vary the number of samples with the physical error rate $p$ to obtain similar error bars for each $p$. We calculate the logical failure rate as $\text{LFR}(p)=1-(1-p_L)^{1/{T}}$ \cite{xu2024constant}, where $p_L$ is the rate at which a logical error occurs at any logical qubit at any round. We report the LFRs for $X$-type logical errors only; $Z$-type logical errors can be simulated similarly.

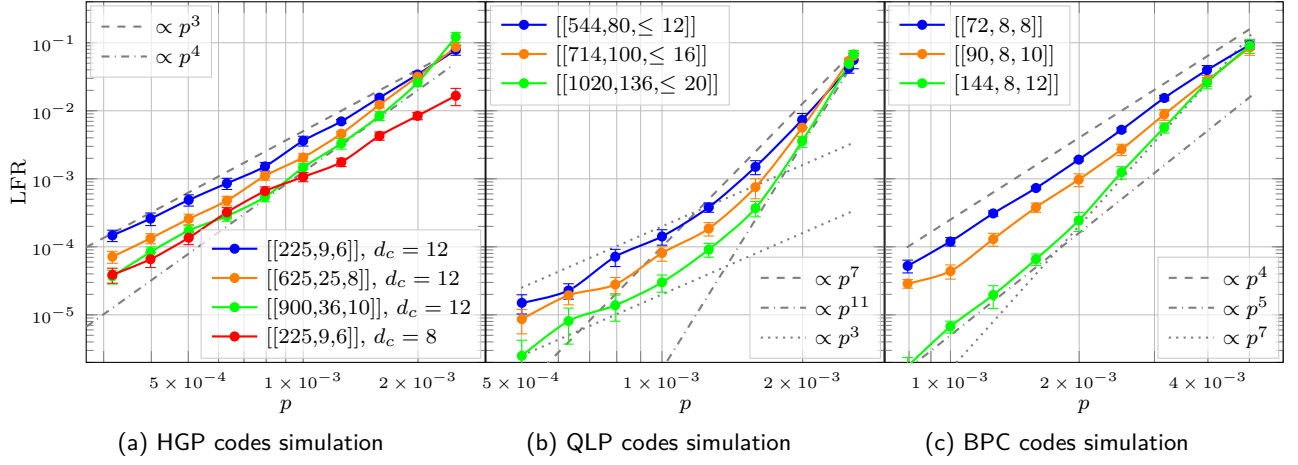
\begin{figure*}[ht!]
\begin{subfigure}[b]{0.3\textwidth}
    \centering
    \input{new_thre_plots/HGP}
    \vspace{-6mm}
    \captionsetup{margin={15mm,0mm}}
    \caption{HGP codes simulation}
    \label{fig:1}
\end{subfigure}
\hspace{7.9mm}
\begin{subfigure}[b]{0.3\textwidth}
    \centering
    \input{new_thre_plots/QLP}
    \vspace{-6mm}
    \caption{QLP codes simulation}
    \label{fig:2}
\end{subfigure}
\hspace{-0.9mm}
\begin{subfigure}[b]{0.3\textwidth}
    \centering
    \input{new_thre_plots/BPC}
    \vspace{-6mm}
    \caption{BPC codes simulation}
    \label{fig:3}
\end{subfigure}
\caption{\textbf{LFR simulation.} Logical memory simulations of various QLDPC codes under the standard circuit-level depolarizing noise model using a $(5,3)$ space-time sliding-window decoding with a BP-OSD inner decoder. The error bars represent the 95\% confidence interval. Reference curves are drawn for the comparison of effective distances $d_{\rm eff}$. \textbf{(a)} HGP code simulation. A threshold is identified at approximately $ 0.23\%$ for circuits with syndrome extraction depth $d_c=12$. Note that the $[[225,9,6]]$ HGP code with $d_c = 8$ outperforms all HGP codes implemented with a $d_c = 12$ circuit in the high-error region and exhibits comparable performance to the $[[900,36,10]]$ code with a $d_c = 12$ circuit as $p$ decreases. \textbf{(b)} QLP code simulation. A threshold is identified at approximately $0.24\%$. Notably, an error floor behavior is observed for the QLP code. This is evident as the slopes of the curves reduce, showing no clear dependence on the code size. This error floor is likely caused by small cycles in the space-time detector error matrix, which hinder the convergence of BP. \textbf{(c)} BPC code simulation. A threshold is observed at approximately $0.47\%$. The slopes of the curves suggest that our circuits maintain $d_{\rm eff}$ comparable to $d$ for the simulated range of $p$.}
\label{fig:threshold}
\end{figure*}

\subsection{Threshold plots}

We present our results in Fig.~\ref{fig:threshold}. All simulations are performed using $(W=5,\:F=3)$ sliding-window decoding with the BP-OSD decoder~\cite{roffe2020decoding} as the inner decoder. We set the BP-OSD parameters as $(\mathcal{I}, \: O) = (10, 1)$ for the HGP and QLP codes and $(20, 10)$ for the BPC codes, where $\mathcal{I}$ is the maximum number of BP iterations and $O$ is the OSD order. We expect the results can be further improved by optimizing the decoding parameters; see Sec.~\ref{subsec:tradeoff} for a detailed discussion. 

In Fig.~\ref{fig:threshold}(a), we plot the LFRs of all three HGP codes implemented with syndrome extraction circuits of $d_c=12$. A threshold is identified at approximately $0.23\%$. Comparing our results for HGP625 with those in Ref.~\cite{tremblay2022constant}, which uses the single-shot decoder based on BP and small-set flip~\cite{grospellier2021combining}, the LFR values of our simulations are approximately an order of magnitude lower, whereas the threshold value is on par. 

In addition to the threshold simulation, the red curve in Fig.~\ref{fig:threshold}(a) shows the results for the HGP225 code with a $d_c=8$ circuit, found by the method described in Sec.~\ref{sec:syndcircuits}. This circuit saturates the lower bound of $d_c$ derived in Ref.~\cite{tremblay2022constant} for this family of codes; see Appx.~\ref{app:depth}. HGP225 with $d_c=8$ outperforms all three HGP codes with $d_c=12$ near the threshold, even though HGP625 and HGP900 have higher code distances. Even for $p \approx 3\times 10^{-4}$, the performance of HGP225 with $d_c=8$ is on par with that of HGP900 with $d_c=12$. This significant improvement in the logical performance arises from the fact that fewer error locations exist in the circuit with lower $d_c$.

The slope of the error-rate curve represents $\lfloor (d_{\rm eff}+1)/2\rfloor$, where $d_{\rm eff}$ is the \textit{effective distance} of the code, which indicates that the code can correct up to approximately $(d_{\rm eff}-1)/2$ independent errors. Note that here we do not consider methods that use post-selection to discard shots for which two minimum-weight predictions of physical errors have the same weight and correspond to different logical classes, which is common in even-distance codes~\cite{reichardt2024demonstration, paetznick2024demonstration}. Before analyzing the results, we briefly explain how the slope may depend on the error rate $p$. For an ideal decoder that is guaranteed to correct all error configurations of weight $\lfloor (d-1)/2\rfloor$, such as the minimum-weight perfect matching decoder of the surface code, the slope is $\lfloor (d+1)/2\rfloor$ for $\forall p$ that is sufficiently below the threshold. However, more heuristic decoders such as BP may fail to decode a few specific error configurations of relatively low weight, especially in the circuit level. The number of error configurations of weight $w$ that lead to a logical failure is denoted as $A(w)$, where $A(w)$ typically increases with $w$. Then, the probability of logical failure due to an error of weight $w$ is given by $A(w)p^w$. At relatively high $p$, the total LFR may be dominated by $A(w)p^w$ of larger $w$, due to the larger value of $A(w)$. However, as $p$ is decreased, $A(w)p^w$ of a smaller $w$ value eventually dominates, as $p^w$ is larger for smaller $w$. Thus, as $p$ is decreased, the slope $\lfloor (d_{\rm eff}+1)/2\rfloor$ saturates to a lower value, which represents the minimum weight of errors that cause a logical failure. A reduction from $d$ to $d_{\rm eff}$ can result from hook errors in syndrome extraction circuits or the decoding process; see Sec.~\ref{sec:outlook} for a detailed discussion and potential strategies to improve $d_{\rm eff}$.

\begin{figure*}[ht!]
  \centering
  \includegraphics[width=\textwidth]{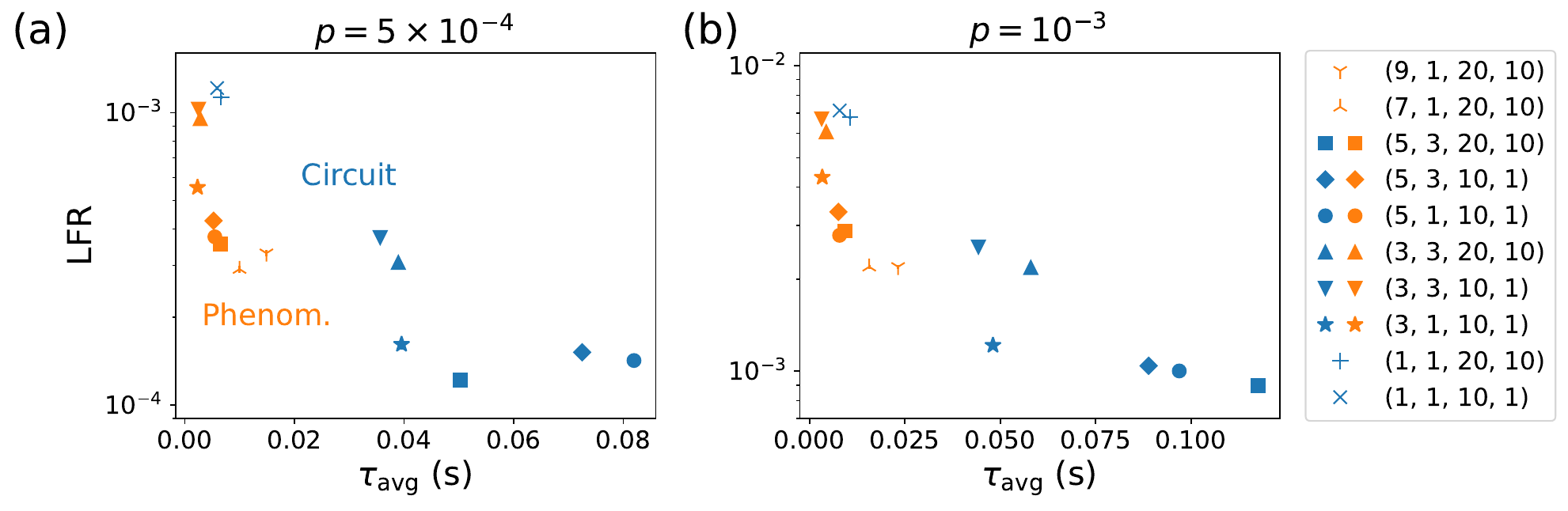}
  \caption{\textbf{Trade-offs between the decoding runtime and the LFR.} Logical memory of the HGP225 code with $d_c=8$ circuit is simulated using various decoder parameters under the standard circuit-level depolarizing noise model where \textbf{(a)} $p=5\times 10^{-4}$ and \textbf{(b)} $p=10^{-3}$. For all points, BP-OSD decoder is used as the inner decoder. We report the LFR for $T=16$ rounds of noisy syndrome extraction circuit and the total decoding runtime divided by the number of times the inner decoder is called. The numbers in the legend represent $(W, F, \mathcal{I}, O)$ where $W$ is the width of the sliding window, $F$ is the offset between consecutive sliding windows, $\mathcal{I}$ is the maximum number of BP iterations, and $O$ is the OSD order. For blue symbols, detector error matrices for circuit-level noise are used. For orange symbols, detector error matrices for the phenomenological noise model are used. The runtimes, averaged over 20000 samples for each parameter set, are evaluated in a single node of the Duke Computing Cluster. Error bars are smaller than the symbols and thus not shown.}
  \label{fig:tradeoff}
\end{figure*}

For an easy comparison of $d_{\rm eff}$, in Fig.~\ref{fig:threshold}(a), we show two reference curves in gray. The slope of the HGP225 curve is near $3$, which indicates that our circuit simulation achieves \mbox{$d_{\rm eff} \approx d$} for HGP225 in the simulated region of $p$. For HGP900, $d_{\rm eff} \approx 7 < d$ near the threshold and is further reduced for lower values of $p$.

Figure~\ref{fig:threshold}(b) shows the results for the QLP codes. The codes we simulate here are introduced in Ref.~\cite{xu2024constant}. The threshold is identified at approximately $0.24\%$. In the low-$p$ region, all QLP codes display comparable slopes, suggesting the reduction of $d_{\rm eff}$ similarly to the HGP900 code. The slopes indicate that the lowest weight of uncorrectable errors in these codes is approximately the same, despite the different code distances $d$. Such error floor behavior is likely due to the difficulty of the BP-based decoder in handling small cycles in the detector error matrix. Similar behavior is observed for the simulation results in Ref.~\cite{xu2024constant}.

Figure~\ref{fig:threshold}(c) shows the results for the BPC codes introduced in Ref.~\cite{tiew2024low}. This family of codes exhibits a relatively high circuit-level threshold at approximately $0.47\%$ due to the low degree of connectivity. The slopes of the curves indicate that \mbox{$d_{\rm eff} \approx d$} for the simulated range of $p$, although error floors may still occur in the region of lower $p$ for BP-based decoders.

\subsection{Trade-offs between time and accuracy}\label{subsec:tradeoff}

Even after the QEC code, the syndrome extraction circuit, and the inner decoder are determined, many degrees of freedom remain for implementing the decoder. In this section, we discuss the trade-offs between the decoding runtime and the LFR that are relevant for choosing the appropriate set of decoder parameters. 

Figure~\ref{fig:tradeoff} shows the results for simulating the logical memory circuit of the HGP225 code with syndrome extraction depth $d_c=8$. The average decoding runtime $\tau_{\rm avg}$ is given by $\tau/N$, where $\tau$ is the total runtime for decoding $T$ rounds and $N$ is the number of times the inner decoder is called. Note that we use the python version of the BP-OSD decoder~\cite{roffe2020decoding} provided by the LDPC package~\cite{Roffe_LDPC_Python_tools_2022} as the inner decoder; we expect that $\tau_{\rm avg}$ can be overall reduced by using the C\texttt{++} version~\cite{Roffe_LDPC_Python_tools_2022}. Our goal here is to compare the trend of LFR with respect to $\tau_{\rm avg}$ over various sets of parameters for the sliding window and the BP-OSD decoder. 

We observe that there is an overall trend of decreasing LFR as $\tau_{\rm avg}$ increases. This highlights that when evaluating the performance of a decoder, the accuracy and the runtime need to be reported together. For example, while the LFR can be further reduced by increasing the window size $W$ with fixed $F$, the detector error matrix would become too large for realistic implementation, especially for larger codes. Also, the optimal set of decoder parameters within a runtime constraint may depend on the physical error rate $p$. We anticipate that \texttt{QUITS} leads to a benchmark for a fair comparison among various decoders. 

Among the decoder parameters considered here, increasing $W$ most significantly reduces the LFR at the cost of larger $\tau_{\rm avg}$. The decoding runtime increases with $W$ due to the larger size of the detector error matrix slice for each decoding window. The LFR decreases with $W$, as a larger decoding window provides more syndrome information to better distinguish space-like and time-like errors that occur at the beginning of the window. 

Improving both $\mathcal{I}$ and $O$ is necessary for achieving the lowest LFR; however, the overall performance improvement is minor compared to the effects of increasing $W$. Comparing the blue and orange symbols, using detector error matrices for circuit-level noise is essential for achieving low LFR, as explained in Sec.~\ref{sec:decoder}. Interestingly, with suitable choices of $(W, F, \mathcal{I}, O)$, using detector error matrices for the phenomenological noise model can achieve a similar LFR with a lower $\tau_{avg}$ compared to using detector error matrices for circuit-level noise. Future work may explore the trade-offs for other decoder options, such as using parallelization for BP.

\section{Conclusion and Outlook}\label{sec:outlook}

In this work, we introduced \QUITS{}, an open-source, modular framework for circuit-level simulation of quantum low-density parity-check (QLDPC) codes. \QUITS{} enables end-to-end simulations by allowing users to flexibly combine various QEC codes, syndrome extraction circuits, decoding algorithms, and noise models. As illustrated in Fig.~\ref{fig:QUITS}, the modular design of \QUITS{} facilitates performance comparisons between different implementations by allowing users to easily switch or replace individual components, including the QLDPC code, the syndrome extraction circuit, the decoder, and the decoding window configuration. 

Our framework supports several prominent QLDPC families, including hypergraph product (HGP) codes, quasi-cyclic lifted product (QLP) codes, and balanced product cyclic (BPC) codes. Thanks to its modular structure, \QUITS{} can be readily extended to incorporate new code families and decoding algorithms. As part of the framework, we introduce a syndrome extraction circuit that applies across all three code families. Notably, for the HGP225 code, our circuit achieves a lower syndrome extraction depth than the original cardinal circuit in Ref.~\cite{tremblay2022constant}, resulting in smaller detector error matrices and improved logical failure rates.

By applying \QUITS{} to simulate the circuit-level performance of various QLDPC codes, we observe a notable gap between the effective distance $d_{\mathrm{eff}}$ and the original code distance $d$ in the low-error-rate regime, as shown in Sec.~\ref{sec:results}. This observation points to several directions for future investigation.

First, one potential cause of the reduced effective distance is the suboptimal performance of belief propagation (BP) decoding within the sliding-window framework. Specifically, when BP is applied to the Tanner graph of the detector error matrix, its performance can be degraded by certain subgraph structures known as trapping sets~\cite{richardson2003error}, which are known to cause decoding failures and give rise to error floors. One future direction is to investigate these failure mechanisms using \QUITS{} by identifying and analyzing the trapping sets responsible for decoding breakdowns in circuit-level simulations. This may further guide our strategies to mitigate error floors and improve the effective distance, either by improving the BP decoding algorithm itself~\cite{kasai2025efficient} or by finding a better Tanner graph of the detector error matrix. 

Second, hook errors in the syndrome extraction circuit may also cause a reduction of the effective distance. If the high-weight hook errors overlap with the support of a logical observable, a logical error may occur from error mechanisms of weight lower than $\lfloor (d-1)/2 \rfloor$, resulting in the reduction of $d_{\rm eff}$. For HGP codes, Refs.~\cite{quintavalle2022reshape, manes2025distance} prove that hook errors do not reduce $d_{\rm eff}$ in the cardinal circuit~\cite{tremblay2022constant}; however, this proof does not apply to other QLDPC codes. A design principle for syndrome extraction circuits of QLP and BPC codes that suppresses hook errors and maximizes $d_{\rm eff}$ is highly desired. The performance of such syndrome extraction circuits can be efficiently tested using our framework. 

Finally, \QUITS{} is currently limited to simulating the logical memory of QLDPC codes. A future direction is to enable simulations of logical operations, such as the joint measurement of logical observables~\cite{xu2024constant, cross2024improved}. The circuit that performs the logical operation may be co-designed with the decoder implementation, such that $d_{\rm eff}$ of the logical operation is not reduced compared to that of the logical memory, as recently demonstrated for the bivariate bicycle codes~\cite{cross2024improved}. We hope that \QUITS{} extends to providing an end-to-end framework for simulating logical operations of various QLDPC codes.

\section{Acknowledgement}

The authors thank Shilin Huang for the valuable discussions. This work was supported by ARO/LPS QCISS program (W911NF-21-1-0005), the NSF QLCI for Robust Quantum Simulation (OMA-2120757), and NSF Grant 2106213. Any opinions, findings, and conclusions or recommendations expressed in this material are those of the authors and do not necessarily reflect the views of the NSF, ARO or LPS.

\bibliographystyle{plainnat}
\bibliography{bib}

\onecolumn\newpage
\appendix

\section{Syndrome extraction depth}\label{app:depth}

The directions of the quantum Tanner graph's edges are determined by assigning the signs of the classical Tanner graph's edges, as explained in Sec.~\ref{sec:syndcircuits}. For each vertex of the classical Tanner graph, the incident edges need to be distributed $-$ and $+$ signs as evenly as possible, such that the syndrome extraction depth of the syndrome extraction circuit is minimized~\cite{tremblay2022constant}. We present a simple heuristic algorithm that balances the signs for each vertex below. 

\begin{algorithm}[h!]
    \caption{\label{algo:sign} Balanced sign assignment}
    \DontPrintSemicolon
    \KwIn{List $E$ of classical Tanner graph's edges $(i,j)$ ($i\in[r]$, $j\in[n]$).}
    \KwOut{Signs $\sigma_{ij}=-1$ or $+1$ assigned to each edge $(i,j)\in E$}
  \vspace*{5pt}
  \hrule
  \vspace*{5pt}
   \textbf{Initialization:} $\mu_i=0$, $\nu_j=0$ $\forall\: i\in [r],\: j\in[n]$\;
    \For{$(i,j) \in E$}{
    Generate a random number $\xi\in[0,1)$\;
    \uIf{$\mu_i+\nu_j>0$ \textbf{or} $\mu_i+\nu_j=0$ \textbf{and} $\xi\geq0.5$}{
        $\sigma_{ij}=+1$, \: $\mu_i = \mu_i - 1$, \: $\nu_j = \nu_j - 1$\;
    }
    \Else{
        $\sigma_{ij}=-1$, \: $\mu_i = \mu_i + 1$, \: $\nu_j = \nu_j + 1$\;
    }
    }
    \Return $\{\sigma_{ij}\}$
    \label{alg:heurbpssf}
\end{algorithm}

We compare the syndrome extraction depths of the syndrome extraction circuits generated by our method to those of the methods in Refs.~\cite{tremblay2022constant, xu2024constant}. In our implementation of Ref.~\cite{tremblay2022constant}, to assign the signs of the classical Tanner graph's edges, a random permutation $\mathcal{P}$ is applied to the vertices $[n+r]$, such that $\mathcal{P}(j)$ and $\mathcal{P}(n+i)$ represent the positions of the $j$th data bit and $i$th check bit, respectively ($j \in [n]$, $i \in [r]$).  Then, for each edge $(i,j)$ of the Tanner graph, $\sigma_{ij}=+1$ if $\left((\mathcal{P}(n+i) - \mathcal{P}(j)) \: \text{mod}\: (n+r) \right)\leq (n+r)/2$, else $\sigma_{ij}=-1$. The directions of the quantum Tanner graph's edges are assigned the same way as described in Sec.~\ref{sec:syndcircuits}. Note that we assume a random permutation is applied, as an explicit method for finding a balanced ordering is not provided in Ref.~\cite{tremblay2022constant}. We do not report the BPC code circuit's syndrome extraction depth for this method, as the positions of the classical vertices are ill-defined. Meanwhile, in Ref.~\cite{xu2024constant}, the $Z$-type and $X$-type syndromes are separately extracted. The syndrome extraction depth is deterministically given by $4\Delta_c$, where $\Delta_c$ is the maximum degree of the check node in the classical Tanner graph. We note that this syndrome extraction circuit is designed to suit the parallel rearrangement capabilities of the reconfigurable atom array platform~\cite{bluvstein2024logical}, and a method for lowering the syndrome extraction depth is suggested in Ref.~\cite{xu2024constant}. 

\begin{table}[ht!]
\centering
\caption{\label{tab:depth} Minimum syndrome extraction depths of the QLDPC codes used in this work obtained by various methods. The probability of obtaining the minimum depth, out of 1000 random instances, is given inside the parentheses.}
\begin{tabular}{c|c|c|c|c|c}
& HGP225 & HGP625 & HGP900 & QLP & BPC  \\
\hline                      
M. A. Tremblay et al. \cite{tremblay2022constant} & 12 (36.6\%) & 12 (12.9\%) & 12 (7.9\%) & 12 (38.3\%) & - \\
\hline
Q. Xu et al. \cite{xu2024constant} & 16 & 16 & 16 & 20 & 12 \\
\hline
\textbf{This work} & \textbf{8} (1.0\%) & 12 (100\%)  & 12 (100\%) & 12 (75.4\%) & \textbf{8} (100\%)
\end{tabular}
\end{table}

Table~\ref{tab:depth} shows the results. For all QLDPC codes considered in this work, our method either reduces the minimum syndrome extraction depth or improves the probability of obtaining the circuit with minimum syndrome extraction depth. This shows the advantage of Algorithm~\ref{algo:sign} over the previous method in which the relative positions of the bits determine the signs of the classical Tanner graph's edges. 

Notably, for the $[[225, 9, 6]]$ HGP code, circuits with syndrome extraction depth 8 are found. This achieves the lower bound on the syndrome extraction depth, which is simply the degree of the HGP code's Tanner graph~\cite{tremblay2022constant}. A sufficient condition for meeting this lower bound is that the Tanner graphs of the two underlying classical codes consist only of vertices with even degrees and allow for a balanced ordering~\cite{tremblay2022constant}. Our circuit shows that this condition is not a necessary condition, as each data bit is supported by 3 checks in the underlying classical code of this HGP code.

\section{Circuit distance of BPC codes}\label{app:dist}

The reduction in the effective distance of logical performance, as observed in Sec.~\ref{sec:results}, may occur from (i) hook errors in the syndrome extraction circuit and (ii) suboptimal decoding, due to failure of BP and/or insufficient sliding-window width. As \texttt{QUITS} offers a flexible framework for exploring different inner decoders and sliding-window parameters, here we isolate the impact of hook errors by analyzing the \textit{circuit distance}, the minimum weight of error mechanisms in the circuit that can flip a logical observable. Note that while the circuit distance is proven to be equal to the original code distance for the HGP code~\cite{manes2025distance}, the reduction in circuit distance is reported for bivariate bicycle codes~\cite{bravyi2024high}. 

For various BPC codes, the circuit distance is estimated using the Stim function \texttt{search\_for\_undetectable\_logical\_errors}~\cite{gidney2021stim}. This function performs a heuristic search for combinations of error mechanisms that, when combined, flip a logical observable without triggering any detector. The search ends when such a combination is found. An upper bound on the circuit distance is given by the number of error mechanisms in this combination. The scripts and parameters for the search can be found in \QUITS{} documentation. We choose the number of syndrome extraction rounds of the circuit such that the memory required for the search is reasonable for each BPC code; for example, finding the circuit distance for 3 rounds of $[[144, 8, 12]]$ code was infeasible with 100 GB of memory. 

\begin{table}[ht!]
\centering
\caption{\label{tab:dist} Upper bounds on the circuit distances of BPC codes' syndrome extraction circuits}

\begin{tabular}{c|c|c}
    BPC code parameters & Number of rounds & Circuit distance \\
    \hline
     $[[72,8,8]]$ & 3 & $\leq 6$ \\
     \hline
     $[[90,8,10]]$ & 3 & $\leq 8$ \\
     \hline
     $[[144,8,12]]$ & 2 & $\leq 9$ 
\end{tabular}
\end{table}

The results are presented in Table~\ref{tab:dist}. The reduction of the upper bound on the circuit distances from the original code distance indicates that hook errors potentially contribute to the reduced slopes we observe for BPC codes in Fig.~\ref{fig:threshold}(c). Future work needs to be done on optimizing the syndrome extraction circuits such that hook errors are minimized.

\end{document}

%% file: new_thre_plots/HGP.tex
\definecolor{darkgreen}{RGB}{0,100,0}
\definecolor{lightgray204}{RGB}{204,204,204}

\begin{tikzpicture}[scale=1.0]
\begin{axis}[
  trim axis left,
  trim axis right,
  width=2.7in,
  height=2.5in,
  ymode=log,
  xmode=log,
  grid=major,
  ymin=2e-6, ymax=4e-1,
  xlabel={$p$},
  xlabel style={
      font=\scriptsize,
      at={(axis description cs:0.5,0)}, 
      yshift=1ex                        
  },
  ylabel={LFR},
  ylabel style={
      font=\scriptsize,
      at={(axis description cs:0,0.5)}, 
      yshift=-2ex                        
  },
  ticklabel style = {font=\footnotesize},
  xmin=0.00027, xmax=0.003,
  xtick={0.0001,0.0002,0.0003,0.0004,0.0005,0.0006,0.0007,0.0008,0.0009,0.001,0.002,0.003},
  xticklabels={$1\times10^{-4}$,,,,$5\times10^{-4}$,,,,,$1\times10^{-3}$,$2\times10^{-3}$,},
  xticklabel style={font=\tiny},
  yticklabel style={font=\tiny},
]

\addplot[
    domain=0.00027:0.0025,
    dashed,
    thick,
    color=gray,
    opacity=1.0,
    samples=200
]
{ x^3 * 10^(6.7) };
\label{$propto3$}

\addplot[
    domain=0.00027:0.0025,
    dashdotted,
    thick,
    color=gray,
    opacity=1.0,
    samples=200
]
{ x^4 * 10^(9.1) };
\label{$propto4$}

\addplot+[
  solid, thick, blue, mark=*, mark options={scale=0.75},
  smooth, 
  error bars/.cd, 
    y dir=both, 
    y explicit
] table [x=x, y=y, y error plus=error, y error minus=error, col sep=comma] {
  x,           y,         error
0.0003162, 0.000147384, 2.81E-05
0.0003981, 0.000259433, 5.45E-05
0.0005012, 0.000488392, 9.17E-05
0.000631, 0.00085705, 0.00016083
0.0007943, 0.00151714, 0.000221476
0.001, 0.00362173, 0.00059681
0.0012589, 0.00697683, 0.000477282
0.0015849, 0.0155884, 0.00137446
0.0019953, 0.0341267, 0.00360564
0.0025119, 0.0785745, 0.0131238
};
\label{[[225,9,6]], $d_c = 12$}

\addplot+[
  solid, thick, orange, mark=*, mark options={scale=0.75},
  smooth, 
  error bars/.cd, 
    y dir=both, 
    y explicit
] table [x=x, y=y, y error plus=error, y error minus=error, col sep=comma] {
  x,           y,         error
0.0003162, 7.15E-05, 1.43E-05
0.0003981, 0.000133069, 2.25E-05
0.0005012, 0.000256744, 4.54E-05
0.000631, 0.000476701, 7.58E-05
0.0007943, 0.00110282, 0.00014908
0.001, 0.00205826, 0.000267494
0.0012589, 0.00460908, 0.000360024
0.0015849, 0.0122582, 0.00120492
0.0019953, 0.0316751, 0.00344212
0.0025119, 0.0865354, 0.0142761
};
\label{[[625,25,8]], $d_c = 12$}

\addplot+[
  solid, thick, green, mark=*, mark options={scale=0.75},
  smooth, 
  error bars/.cd, 
    y dir=both, 
    y explicit
] table [x=x, y=y, y error plus=error, y error minus=error, col sep=comma] {
  x,           y,         error
0.0003162, 3.68E-05, 8.44E-06
0.0003981, 8.39E-05, 1.57E-05
0.0005012, 0.000172843, 3.15E-05
0.000631, 0.000283806, 4.62E-05
0.0007943, 0.000539679, 8.06E-05
0.001, 0.00146236, 0.000225032
0.0012589, 0.00331009, 0.000581722
0.0015849, 0.00841225, 0.00123503
0.0019953, 0.0259823, 0.00305313
0.0025119, 0.121874, 0.0201246
};
\label{[[900,36,10]], $d_c = 12$}

\addplot+[
  solid, thick, red, mark=*, mark options={scale=0.75},
  smooth, 
  error bars/.cd, 
    y dir=both, 
    y explicit
] table [x=x, y=y, y error plus=error, y error minus=error, col sep=comma] {
  x,           y,         error
0.0003162, 3.88E-05, 9.65E-06
0.0003981, 6.57E-05, 1.62E-05
0.0005012, 0.000136076, 2.86E-05
0.000631, 0.000320771, 5.56E-05
0.0007943, 0.000657398, 0.0001028
0.001, 0.00106471, 0.000160442
0.0012589, 0.00175506, 0.000246762
0.0015849, 0.00428055, 0.000593584
0.0019953, 0.00841732, 0.00103363
0.0025119, 0.0166031, 0.00472121
};
\label{[[225,9,6]], $d_c = 8$}

\node[
  fill=white,
  fill opacity=0.7,
  draw opacity=1,
  text opacity=1,
  anchor=north west,
  draw=lightgray204,
  font=\scriptsize,
] at (rel axis cs: 0.01,0.99) {
\shortstack[l]{
\ref{$propto3$} $\propto p^3$\\
\ref{$propto4$} $\propto p^4$
}
};

\node[
  fill=white,
  fill opacity=0.7,
  draw opacity=1,
  text opacity=1,
  anchor=south east,
  draw=lightgray204,
  font=\scriptsize,
] at (rel axis cs: 0.99,0.01) {
\shortstack[l]{
\ref{[[225,9,6]], $d_c = 12$} [[225,9,6]], $d_c = 12$\\
\ref{[[625,25,8]], $d_c = 12$} [[625,25,8]], $d_c = 12$\\
\ref{[[900,36,10]], $d_c = 12$} [[900,36,10]], $d_c = 12$\\
\ref{[[225,9,6]], $d_c = 8$} [[225,9,6]], $d_c = 8$
}
};

\end{axis}
\end{tikzpicture}

%% file: new_thre_plots/QLP.tex
\definecolor{darkgreen}{RGB}{0,100,0}
\definecolor{lightgray204}{RGB}{204,204,204}

\begin{tikzpicture}[scale=1.0]
\begin{axis}[
  trim axis left,
  trim axis right,
  width=2.7in,
  height=2.5in,
  ymode=log,
  xmode=log,
  grid=major,
  ymin=2e-6, ymax=4e-1,
  xlabel={$p$},
  xlabel style={
      font=\scriptsize,
      at={(axis description cs:0.5,0)}, 
      yshift=1ex                        
  },
  ticklabel style = {font=\footnotesize},
  xmin=0.00042, xmax=0.003,
  xtick={0.0001,0.0002,0.0003,0.0004,0.0005,0.0006,0.0007,0.0008,0.0009,0.001,0.002,0.003},
  xticklabels={$1\times10^{-4}$,,,,$5\times10^{-4}$,,,,,$1\times10^{-3}$,$2\times10^{-3}$,},
  xticklabel style={font=\tiny},
  yticklabels={},
  yticklabel style={font=\tiny},
]

\addplot[
    domain=0.0005:0.0026,
    dashed,
    thick,
    color=gray,
    opacity=1.0,
    samples=200
]
{ x^7 * 10^(17) };
\label{propto_7}

\addplot[
    domain=0.0005:0.0026,
    dashdotted,
    thick,
    color=gray,
    opacity=1.0,
    samples=200
]
{ x^11 * 10^(27.2) };
\label{propto11}

\addplot[
    domain=0.0005:0.0026,
    dotted,
    thick,
    color=gray,
    opacity=1.0,
    samples=200
]
{ x^3 * 10^(5.3) };
\label{propto3_2}

\addplot[
    domain=0.0005:0.0026,
    dotted,
    thick,
    color=gray,
    opacity=1.0,
    samples=200
]
{ x^3 * 10^(4.3) };
  
  \addplot+[
    solid, thick, blue, mark=*, mark options={scale=0.75},
  smooth, 
  error bars/.cd, 
    y dir=both, 
    y explicit
  ] table [x=x, y=y, y error plus=error, y error minus=error, col sep=comma] {
  x,           y,         error
0.0005012, 1.50E-05, 4.72E-06
0.000631, 2.28E-05, 5.81E-06
0.0007943, 7.15E-05, 2.02E-05
0.001, 0.000141976, 3.62E-05
0.0012589, 0.000380892, 5.94E-05
0.0015849, 0.00149583, 0.000347692
0.0019953, 0.0073732, 0.00176003
0.0025119, 0.0423967, 0.00677269
0.0025704, 0.0556592, 0.0141681
};
\label{544}

  \addplot+[
  thick,
  orange,
  solid,   
  mark=*, 
  mark options={scale=0.75},
  smooth, 
  error bars/.cd, 
    y dir=both, 
    y explicit
] table [x=x, y=y, y error plus=error, y error minus=error, col sep=comma] {
  x,           y,         error
0.0005012, 8.63E-06, 3.38E-06
0.000631, 1.94E-05, 5.38E-06
0.0007943, 2.79E-05, 7.44E-06
0.001, 8.09E-05, 1.95E-05
0.0012589, 0.000185351, 4.14E-05
0.0015849, 0.000754252, 0.000246297
0.0019953, 0.00564988, 0.00153183
0.0025119, 0.0539609, 0.00565599
0.0025704, 0.0640675, 0.00908723
};
\label{714}

  \addplot+[
  thick,
  green,
  solid,   
  mark=*, 
  mark options={scale=0.75},
  smooth, 
  error bars/.cd, 
    y dir=both, 
    y explicit
] table [x=x, y=y, y error plus=error, y error minus=error, col sep=comma] {
  x,           y,         error
0.0005012, 2.50E-06, 1.73E-06
0.000631, 8.13E-06, 4.42E-06
0.0007943, 1.38E-05, 5.75E-06
0.001, 3.00E-05, 8.49E-06
0.0012589, 9.13E-05, 2.09E-05
0.0015849, 0.000369774, 9.43E-05
0.0019953, 0.00356236, 0.000661632
0.0025119, 0.0486819, 0.00743835
0.0025704, 0.0675371, 0.00946825
};
\label{1020}

\node [
    fill=white,
    fill opacity=0.7,
    draw opacity=1,
    text opacity=1,
    anchor=north west,
    draw=lightgray204,
    font=\scriptsize,
] at (rel axis cs: 0.01,0.99) {
\shortstack[l]{
\ref{544} [[544,80,$\leq$ 12]]\\
\ref{714} [[714,100,$\leq$ 16]]\\
\ref{1020} [[1020,136,$\leq$ 20]]
}
};

\node [
    fill=white,
    fill opacity=0.7,
    draw opacity=1,
    text opacity=1,
    anchor=south east,
    draw=lightgray204,
    font=\scriptsize,
] at (rel axis cs: 0.99,0.01) {
\shortstack[l]{
\ref{propto_7} $\propto p^7$\\
\ref{propto11} $\propto p^{11}$\\
\ref{propto3_2} $\propto p^3$
}
};

\end{axis}
\end{tikzpicture}

%% file: new_thre_plots/BPC.tex
\definecolor{darkgreen}{RGB}{0,100,0}
\definecolor{lightgray204}{RGB}{204,204,204}

\begin{tikzpicture}[scale=1.0]
\begin{axis}[
  trim axis left,
  trim axis right,
  width=2.7in,
  height=2.5in,
  ymode=log,
  xmode=log,
  grid=major,
  ymin=2e-6, ymax=4e-1,
  xlabel={$p$},
  xlabel style={
      font=\scriptsize,
      at={(axis description cs:0.5,0)}, 
      yshift=1ex                        
  },
  ticklabel style = {font=\footnotesize},
  xmin=0.0007, xmax=0.006,
  xtick={0.0001,0.0002,0.0003,0.0004,0.0005,0.0006,0.0007,0.0008,0.0009,0.001,0.002,0.003,0.004,0.005,0.006},
  xticklabels={$1\times10^{-4}$,,,,$5\times10^{-4}$,,,,,$1\times10^{-3}$,$2\times10^{-3}$,,$4\times10^{-3}$,,},
  xticklabel style={font=\tiny},
  yticklabels={},
  yticklabel style={font=\tiny},
]

\addplot[
    domain=0.00079:0.0051, 
    dashed,
    thick,
    color=gray,
    opacity=1.0,
    samples=200
]
{ x^4 * 10^(8.4) };
\label{propto4_2}


\addplot[
    domain=0.00079:0.0051,
    dashdotted,
    thick,
    color=gray,
    opacity=1.0,
    samples=200
]
{ x^5 * 10^(9.7) };
\label{propto5}

\addplot[
    domain=0.00079:0.0051,
    dotted,
    thick,
    color=gray,
    opacity=1.0,
    samples=200
]
{ x^7 * 10^(15.2) };
\label{propto7}

  \addplot+[
    solid, thick, blue, mark=*, mark options={scale=0.75},
  smooth, 
  error bars/.cd, 
    y dir=both, 
    y explicit
  ] table [x=x, y=y, y error plus=error, y error minus=error, col sep=comma] {
  x,           y,         error
0.0007943, 5.25E-05, 1.12E-05
0.001, 0.000119482, 1.69E-05
0.0012589, 0.000309467, 2.73E-05
0.0015849, 0.0007315, 4.20E-05
0.0019953, 0.00191475, 9.65E-05
0.0025119, 0.00525188, 0.000361285
0.0031623, 0.0154089, 0.00143232
0.0039811, 0.039663, 0.00648212
0.0050119, 0.0929344, 0.0124428
};
\label{$[[72,8,8]]$}

  \addplot+[
    solid, thick, orange, mark=*, mark options={scale=0.75},
  smooth, 
  error bars/.cd, 
    y dir=both, 
    y explicit
  ] table [x=x, y=y, y error plus=error, y error minus=error, col sep=comma] {
  x,           y,         error
0.0007943, 2.88E-05, 4.16E-06
0.001, 4.38E-05, 1.03E-05
0.0012589, 0.000129814, 2.79E-05
0.0015849, 0.000380892, 5.94E-05
0.0019953, 0.000972974, 0.000206748
0.0025119, 0.00270416, 0.000514139
0.0031623, 0.00882456, 0.00157984
0.0039811, 0.0278249, 0.00519417
0.0050119, 0.0853326, 0.0199389
};
\label{$[[90,8,10]]$}

  \addplot+[
    solid, thick, green, mark=*, mark options={scale=0.75},
  smooth, 
  error bars/.cd, 
    y dir=both, 
    y explicit
  ] table [x=x, y=y, y error plus=error, y error minus=error, col sep=comma] {
  x,           y,         error
0.0007943, 1.76E-06, 5.90E-07
0.001, 6.73E-06, 1.33E-06
0.0012589, 1.96E-05, 7.19E-06
0.0015849, 6.57E-05, 1.28E-05
0.0019953, 0.000244197, 7.66E-05
0.0025119, 0.00124772, 0.000259068
0.0031623, 0.00565441, 0.000969219
0.0039811, 0.0259437, 0.00498133
0.0050119, 0.0908123, 0.0210943
};
\label{$[144,8,12]]$}

\node [
    fill=white,
    fill opacity=0.7,
    draw opacity=1,
    text opacity=1,
    anchor=south east,
    draw=lightgray204,
    font=\scriptsize,
] at (rel axis cs: 0.99,0.01) {
\shortstack[l]{
\ref{propto4_2} $\propto p^4$\\
\ref{propto5} $\propto p^5$\\
\ref{propto7} $\propto p^7$
}
};

\node [
    fill=white,
    fill opacity=0.7,
    draw opacity=1,
    text opacity=1,
    anchor=north west,
    draw=lightgray204,
    font=\scriptsize,
] at (rel axis cs: 0.01,0.99) {
\shortstack[l]{
\ref{$[[72,8,8]]$} $[[72,8,8]]$\\
\ref{$[[90,8,10]]$} $[[90,8,10]]$\\
\ref{$[144,8,12]]$} $[144,8,12]]$
}
};

\end{axis}
\end{tikzpicture}